\documentclass[final,leqno,onefignum,onetabnum]{siamltex1213}
\usepackage{graphicx}
\usepackage{amssymb}
\usepackage{amsmath}
\usepackage{enumerate}
\usepackage{cite}
\usepackage{mathcomp}
\usepackage{supertabular}
\usepackage{longtable}
\usepackage{stmaryrd}
\usepackage{url}
\usepackage{rotating}
\usepackage{float}
\usepackage{tikz}
\usetikzlibrary{shapes}
\usepackage[ruled,vlined, linesnumbered]{algorithm2e}
\usepackage{array}
\usepackage{mathrsfs}
\usepackage{placeins}
\usepackage{subfigure}

\renewcommand{\thesection}{\arabic{section}}

\newtheorem{example}{Example}[section]
\newcommand\remove[1]{}

\newcommand{\vA}{{\bf A}}

\newcommand{\vB}{{\bf B}}

\newcommand{\vI}{{\bf I}}

\newcommand{\vJ}{{\bf J}}

\newcommand{\vM}{{\bf M}}
\newcommand{\vN}{{\bf N}}
\newcommand{\tN}{{\widetilde{\bf N}}}
\newcommand{\vU}{{\bf U}}

\newcommand{\va}{{\bf a}}
\newcommand{\vb}{{\bf b}}

\newcommand{\vn}{{\bf n}}
\newcommand{\vu}{{\bf u}}
\newcommand{\vv}{{\bf v}}

\newcommand{\vx}{{\bf x}}

\newcommand{\vj}{{\bf j}}
\newcommand{\vzero}{{\bf 0}}

\newcommand{\C}{\mathcal C}
\newcommand{\D}{\mathcal D}

\newcommand{\U}{\mathcal U}
\newcommand{\V}{\mathcal V}

\newcommand{\enarrow}{e_{\sf NBD}}

\newcommand{\eimpulse}{e_{\sf IMP}}

\newcommand{\FF}{\mathbb F}

\newcommand{\floor}[1]{\left\lfloor{#1}\right\rfloor}

\newcommand{\inprod}[1]{\langle{#1}\rangle}

\newcommand{\beas}{\begin{eqnarray*}} 
\newcommand{\eeas}{\end{eqnarray*}} 

\newcommand{\etal}{\emph{et al.}}
\newcommand{\mbyn}{$(m\times n)$-matrix code}

\newcommand\ip[1]{\langle{#1}\rangle}

\title{Product Construction of Affine Codes\thanks{This paper was presented
    in part at IEEE International Symposium on Information Theory 2014.
    Submitted: 10 July 2014. Accepted: 03 June 2015.}}

\author{%
    Yeow Meng Chee\thanks{Division of Mathematical Sciences, School of Physical and Mathematical Sciences,
    Nanyang Technological University, Singapore (\email{ymchee \_at\_
ntu \_dot\_ edu \_dot\_ sg}, \email{hmkiah \_at\_ ntu \_dot\_ edu \_dot\_ sg})}
    \and
    Han Mao Kiah\footnotemark[2]
    \and
    Punarbasu Purkayastha\thanks CGG Services Pte Ltd, Singapore
    (\email{punarbasu+paper \_at\_ gmail \_dot\_ com})
    \and
    Patrick Sol\'e%
    \thanks{Institut Mines-T\'el\'ecom, T\'el\'ecom ParisTech, CNRS LTCI,
            France
            (\email{patrick.sole \_at\_ telecom-paristech \_dot\_ fr})}
}

\begin{document}
\maketitle
\slugger{sidma}{xxxx}{xx}{x}{x--x}

\begin{abstract}
Binary matrix codes with restricted row and column weights are a desirable
method of coded modulation for power line communication. In this work, we
construct such matrix codes that are obtained as products of affine codes
- cosets of binary linear codes. Additionally, the constructions have the
property that they are systematic. Subsequently, we generalize our
construction to irregular product of affine codes, where  the
component codes are affine codes of different rates.
\end{abstract}



\pagestyle{myheadings}
\thispagestyle{plain}
\markboth{Y. M. CHEE, H. M. KIAH, P. PURKAYASTHA, P. SOL\'E}%
         {PRODUCT CONSTRUCTION OF AFFINE CODES}

\section{Introduction}

Product codes were introduced by Elias \cite{Elias:1954} and subsequently
generalized by Forney \cite{Forney:1966} to concatenated codes. Product
codes are a method of constructing larger codes from smaller codes while
retaining the good rates and good decoding complexity from the smaller
codes. The codewords of a product code can be written as matrices with the
rows belonging to the row component code and the columns belonging to the
column component code. List decoding algorithms have also been studied in
this context in Barg and Z{\'e}mor \cite{Barg:2011} where the min-sum algorithm was shown
to be amenable to list decoding of product codes.

Product codes have been subsequently generalized to yield codes obtained
from product of nonlinear codes
by Amrani \cite{Amrani:2007}, and to multilevel product codes by Zinoviev
\cite{Zinoviev:1976}. Amrani \cite{Amrani:2007} gave the
construction of product codes from component nonlinear codes which are
binary and systematic. The construction guarantees that all the columns of
any codeword belong to the column component code; however only the first
few rows corresponding to the systematic part of the column code are
guaranteed to belong to the row code. 
In the case where one of the component codes is linear, 
Amrani \cite{Amrani:2007} proposed two soft-decision decoding algorithms.
Irregular product codes,
introduced by Alipour \etal \cite{Alipour:2012}, are yet another
generalization of product codes where each row and column code can be
a code of different rate. Irregular product codes were introduced to
address the need for unequal error protection from bursty noise when some
parts of the codeword are more vulnerable to burst errors than others.

In this work we study constructions of systematic nonlinear product
codes which are obtained as products of affine codes -- cosets of linear
codes. In contrast to the work of Amrani \cite{Amrani:2007}, our construction 
guarantees that all the rows belong to the (affine) row code and
all the columns belong to the (affine) column code. One primary motivation
for studying such class of codes arises from a previous study on coded
modulation for power line channels by Chee \etal \cite{Cheeetal:2013c}
that proposed a generalization of the coded modulation scheme of
Vinck \cite{Vinck:2000}.

Chee \etal
\cite{Cheeetal:2013c} showed that binary matrix codes with bounded column
weights, in conjunction with multitone frequency shift keying, can be used
to counter the harsh noise characteristics of the power line channel.
Concatenated codes obtained from the concatenation of constant weight inner
codes with Reed-Solomon outer codes were used to obtain families of
efficiently decodable codes with good rates and good relative distances.
In this work, we continue this line of investigation and introduce
{\em binary} systematic product codes
with the additional restriction that the row and column weights are
bounded. The restriction on the column weight arises from the desire to be
able to detect and correct impulse noise that is present in the power line
channel. The restriction on the row weights allows one to detect and
correct narrowband noise. It is quite evident that product codes obtained
from the product of linear codes do not satisfy these restrictions.
The nonlinear codes studied in this paper are constructed to satisfy these
properties.
The efficient decoding algorithms of product codes are directly applicable
to the constructions presented in this work. As a first step to the
decoding process, we subtract the coset representative that is used in the construction.
The coset representative is explicitly described, as explained in the following sections.

The rest of the paper is organized as follows. In the next section we
introduce the basic definitions and notation that are used throughout
the rest of this paper. Section \ref{sec:affine} discusses the general
construction of $q$-ary systematic codes which are product of affine codes.
Section \ref{sec:affine-example} uses the construction
in Section \ref{sec:affine} to give constructions of {\em binary} product codes with
restricted row and column weights. This is of interest because of its
application to coded modulation for power line channels. In Section
\ref{sec:irregular}, we extend this construction to product codes which can
provide unequal error protection, where different rows and columns belong to 
different row and column codes. This section generalizes the irregular product code
construction of Alipour \etal \cite{Alipour:2012}, where the component codes are linear codes,
to irregular product codes where the component codes
are affine codes.

\section{Notation and Definitions}
\label{sec:notation}

Denote the set of integers $\{1,2,\ldots,,n\}$ by $[n]$ for a positive integer $n$.
Denote the finite field of order $q$ by $\FF_q$.
A {\em  code} $\C$ of {length} $n$ is a subset of $\FF_q^n$, while 
a {\em  linear code} $\C$ of {length} $n$ is a linear subspace of $\FF_q^n$.
The {\em dimension} of a linear code $\C$ is given by the dimension of $\C$ as a linear subspace of $\FF_q^n$. 
Elements of $\C$ are called {\em codewords}. 
Endow the space $\FF_q^n$ with the {\em Hamming distance} metric
and for $\vu\in\FF_q^n$, the {\em Hamming weight} of $\vu$ is the distance of $\vu$ from the all-zero codeword.
A code $\C\subseteq \FF_q^n$ is said to have distance $d$ 
if the (Hamming) distance between any two distinct codewords of $\C$ is at least $d$. 
Moreover, a linear code $\C$ has distance $d$ if the weight of all nonzero codewords in $\C$ is at least $d$. 
We use the notation $[n, k, d]$ to denote a linear code of length $n$,
dimension $k$ and distance $d$.

Let $m$, $n$ be positive integers and let $\FF_q^{m\times n}$ denote the set 
of $m$ by $n$ matrices over $\FF_q$. 
The {\em transpose} of a matrix $\vM$ is denoted by $\vM^T$ and 
we regard the vector $\vu\in\FF_q^n$ as a row vector, or a matrix $\vu$ in $\FF_q^{1\times n}$.
Hence, $\vu^T$ denotes a column vector in $\FF_q^{n\times 1}$. 
In addition, let $\vzero_n$ and $\vj_n$ denote the all-zero and all-one vectors of length $n$ respectively,
while $\vI_n$ and $\vzero_{m\times n}$ denote the $(n\times n)$ identity and $(m\times n)$ all-zero matrix respectively.
We denote the {\em span} of a vector $\vu$ by the notation $\ip{\vu}.$

Let $\C$ be a linear $[n,k,d]$ code. After a permutation of coordinates, 
there exists a matrix $\vA\in \FF_q^{k\times (n-k)}$ such that 
each codeword in $\C$ can be written as $(\vx,\vx\vA)$,
where $\vx\in\FF_q^k$ is called the {\em information vector}.
The matrix $(\vI_k|\vA)$ is said to be a {\em systematic encoder}
of $\C$.

Let $\C_1$ and $\C_2$ be linear $[n,k_1,d_1]$ and $[n,k_2,d_2]$ codes, respectively.
Suppose $\C_1\subseteq \C_2$ and pick $\vu\in\C_2$. 
Then the set of codewords $\C_1+\vu$ is a {\em coset of $\C_1$ in $\C_2$}.
The collection of all cosets of $\C_1$ in $\C_2$ is denoted by $\C_2/\C_1$.
Moreover, any coset in $\C_2/\C_1$ is a $(n,d_1)$ code of size $q^{k_1}$,
and we call the coset an {\em affine} $[n,k_1,d_1]$ code.

Observe that if $(\vI_{k_1}|\vA_1)$ is a systematic encoder for $\C_1$ and 
$\vu=(\vu_1,\vu_2)$ where $\vu_1$ is of length $k_1$, 
then $\C_1+\vu=\C_1+(\vzero_{k_1},\vu_2-\vu_1\vA_1)$.
On the other hand, every coset in $\C_2/\C_1$
contains at most one element of the form $(\vzero_{k_1},\va)$. 
 Hence, for every coset $\C_1+\vu$, there is exactly one element of the
 form $(\vzero_{k_1},\va)$,
and in this paper, we refer to this element as the {\em coset representative of $\C_1+\vu$}. 
The set of all coset representatives of cosets in $\C_2/\C_1$ is denoted
$(\C_2/\C_1)_{\rm rep}$.

We also consider the notion of systematicity for nonlinear codes.
Let $\C$ be a (matrix) code of size $q^k$. 
Then $\C$ is said to be {\em systematic of dimension $k$} 
if there exists $k$ coordinates such that $\C$ 
when restricted to these $k$ coordinates is $\FF_q^k$.
Observe that if $\C$ is a linear $[n,k,d]$ code, 
then any 
affine code $\C+\vu$ is systematic of dimension $k$.

 \subsection{ Matrix Codes}
 
An  {\em \mbyn} $\C$ is a subset of $\FF_q^{m\times n}$,
while a {\em  linear \mbyn} $\C$ is a linear
subspace of $\FF_q^{m\times n}$, when considered as a vector space of
dimension $mn$.
Regarding each matrix in $\FF_q^{m\times n}$ as a vector of length $mn$, 
we have the definitions of Hamming distance, Hamming weight and dimension.
A linear \mbyn{} of dimension $K$ and distance $d$ is denoted by $[m\times n,K,d]$.

\subsection{Classical Product Codes}
The classical product code constructs  matrix codes from  two 
linear codes. 
Given a linear $[n,k,d_\C]$ code $\C$ and a linear $[m,l,d_\D]$ code $\D$,
let $(\vI_k|\vA)$ and $(\vI_l|\vB)$ be their respective systematic
encoders.
The {\em product code}, denoted by $\C\otimes \D$, is then given by the
\mbyn{} (see \cite[p. 568]{MacWilliamsSloane:1977})
\begin{equation*}
\C\otimes \D\triangleq\left\{
\left(
 \renewcommand\arraystretch{2} 
\begin{array}{c|c}
\vM & \vM\vA\\ \hline
\vB^T\vM & \vB^T\vM\vA
\end{array}
\right)
:\vM\in \FF_q^{l\times k}
\right\},
\end{equation*}
where $\vM$ corresponds to the information bits.
It can be shown that $\C\otimes \D$ is a linear $[m\times n, kl, d_\D d_\C]$ code.
Furthermore, $\C\otimes\D$ has the following property that 
depends on the {\em component codes $\C$ and $\D$}.
\medskip

\noindent {\bf Property $(\C,\D)$}.
For every $\vN\in\C\otimes \D$,  
 \begin{enumerate}[(i)]
 \item every row of $\vN$ belongs to $\C$, and
 \item every column of $\vN$ belongs to $\D$.
 \end{enumerate}
 
 \medskip
 
 In this paper, we consider nonlinear component codes.
 Specifically, let $\C'$ be a nonlinear code  of length $n$ and size $q^k$  and 
 $\D'$ be a nonlinear code  of length $m$ and size $q^l$.
 We aim to construct an \mbyn{} $\C'\otimes \D'$ of size $q^{kl}$ 
 such that Property $(\C',\D')$ holds.
This construction differs from the nonlinear product code construction in Amrani
\cite{Amrani:2007} because we guarantee that \emph{all} the rows in every codeword
belong to the row code $\C'$.

\section{Product Codes from Affine Codes}
\label{sec:affine}
 \begin{figure*}
 \begin{equation}\label{eq:def1}
 (\C+\vu)\otimes (\D+\vv)\triangleq\left\{
 \left(
 \renewcommand\arraystretch{2} 
 \begin{array}{c|c}
 \vM & \vM \vA+ \vj^T_l \va\\ \hline
 \vB^T \vM+\vb^T\vj_{k} & (\vB^T \vM+\vb^T\vj_{k})\vA+ \vj^T_{m-l}\va
 \end{array}
 \right):\vM\in\FF_q^{l\times k}\right\}.
 \end{equation}
 
  \begin{equation}\label{eq:def2}
 (\C+\vu)\widetilde\otimes (\D+\vv)\triangleq\left\{
 \left(
 \renewcommand\arraystretch{2} 
 \begin{array}{c|c}
 \vM & \vM \vA+ \vj^T_l\va\\ \hline
 \vB^T \vM+\vb^T\vj_k & \vB^T(\vM \vA+ \vj^T_l\va)+\vb^T\vj_{n-k}
 \end{array}
 \right):\vM\in\FF_q^{l\times k}\right\}.
 \end{equation}
 \hrule
\end{figure*}

In this section, we provide the general construction of systematic matrix
codes that are obtained as products of cosets of linear codes, i.e., as
products of affine codes.
Throughout this section,  let $\C$ and $\D$ be a  linear $[n,k,d_\C]$ and $[m,l,d_\D]$ codes respectively.
We consider affine codes that are obtained as cosets of the codes $\C$ and
$\D$, i.e., they are of the form $\C+\vu$ and $\D+\vv$, respectively,
where $\vu$ and $\vv$ are of lengths $n$ and $m$ respectively.
In particular, we show that if both $\C$ and $\D$ contain the all-one vector,
then there exists an \mbyn, that is systematic of dimension $kl$ with
Property $(\C+\vu, \D+\vv)$.
 
 Let $(\vI_k|\vA)$ and $(\vI_l|\vB)$ be systematic encoders for $\C$ and $\D$ respectively.
 Recall that the set of coset representatives of cosets of $\C_1$ in $\C_2$ is
 denoted by $(\C_2/\C_1)_{\mathrm{rep}}$.
 Without loss of generality, pick $\vu=(\vzero_k,\va)\in (\FF_q^n/\C)_{\rm rep}$ and $\vv=(\vzero_l,\vb)\in (\FF_q^m/\D)_{\rm rep}$.
 Then a typical element in $\C+\vu$ is of the form $(\vx,\vx\vA+\va)$ where $\vx$ is the information vector of length $k$.
 Similarly, a typical element in $\D+\vv$ is of the form $(\vx,\vx\vB+\vb)$ where $\vx$ is the information vector of length $l$.
 
 Define $(\C+\vu)\otimes (\D+\vv)$ to be the \mbyn{} given by \eqref{eq:def1}.
This is obtained by the encoding the first $k$ columns by $\D+\vv$, followed by
encoding all the rows by $\C+\vu$.
We observe that for every $\vN\in (\C+\vu)\otimes (\D+\vv)$,
each row of $\vN$ belongs to $\C+\vu$. 
However, we can guarantee only that the first $k$ columns belong to $\D+\vv$.
 
On the other hand, if we alter the definition given in \eqref{eq:def1} to
be \eqref{eq:def2},
where we encode the first $l$ rows by $\C+\vu$, followed by encoding all
the columns using $\D+\vv$,
 we have that every column of $\vN$ belongs to $\D+\vv$ for each $\vN\in  (\C+\vu) \widetilde\otimes (\D+\vv)$.

 Therefore, the matrix code $(\C+\vu)\otimes (\D+\vv)$ meets our requirements if 
 {
 \small
 \begin{align}
 \label{eq:cond1}
 (\vB^T \vM+\vb^T\vj_{k})\vA+ \vj^T_{m-l}\va &=\vB^T(\vM \vA+
     \vj^T_l\va)+\vb^T\vj_{n-k}, \text{that is,} \notag\\
 \vb^T(\vj_k\vA- \vj_{n-k}) &=(\vB^T\vj_l^T-\vj_{m-l}^T)\va 
 \end{align}
 }

 If \eqref{eq:cond1} holds, then $(\C+\vu)\otimes (\D+\vv)$ (or equivalently, $ (\C+\vu) \widetilde\otimes (\D+\vv)$)
 is a coset of $\C\otimes \D$. That is, $(\C+\vu)\otimes (\D+\vv)=(\C\otimes\D)+\vU$, where
 \begin{equation}\label{eq:cl}
 \vU\triangleq\left(
 \renewcommand\arraystretch{2} 
 \begin{array}{c|c}
 {\vzero_{l\times k}} & \vj^T_l\va\\ \hline
 \vb^T\vj_k & \vb^T\vj_{k}\vA+ \vj^T_{m-l}\va
 \end{array}
 \right).
 \end{equation}

 \begin{theorem}\label{prop:product}
Let $\C$ and $\D$ be  linear $[n,k,d_\C]$ and $[m,l,d_\D]$ codes respectively
and $(\vI_k|\vA)$ and $(\vI_l|\vB)$ be their respective systematic encoders.
Pick $\vu=(\vzero_k,\va)\in (\FF_q^n/\C)_{\rm rep}$ and $\vv=(\vzero_l,\vb)\in (\FF_q^m/\D)_{\rm rep}$.

\noindent If in addition \eqref{eq:cond1} holds, then  $(\C+\vu)\otimes (\D+\vv)$ defined by \eqref{eq:def1}
is equal to  $(\C+\vu)\widetilde\otimes (\D+\vv)$ defined by \eqref{eq:def2}.
Moreover, the code is systematic of
dimension $kl$ and is a coset of $\C\otimes \D$
with Property $(\C+\vu,\D+\vv)$.
\end{theorem}
 
We now provide a sufficient condition for \eqref{eq:cond1} to hold.
Observe that $\vj_n\in \C$ if and only if $\vj_k\vA=\vj_{n-k}$, since
$\vj_k(\vI_k | \vA)$ is necessarily $\vj_n$. This is because the only
message vector that can give rise to the all-one vector must have all-one
in the systematic part of the codeword.
Hence, $\vj_k\vA- \vj_{n-k}=\vzero_{n-k}$ and $\vb^T(\vj_k\vA- \vj_{n-k}) =\vzero_{(m-l)\times(n-k)}$. 
Similar argument holds for $\vB^T\vj_l^T-\vj_{m-l}^T$. Hence, \eqref{eq:cond1} holds
and the coset representative $\vU$ given by \eqref{eq:cl} is 

 \begin{equation}
 \vU=\left(
 \renewcommand\arraystretch{2} 
 \begin{array}{c|c}
 {\vzero_{l\times k}} & \vj^T_l\va\\ \hline
 \vb^T\vj_k & \vb^T\vj_{n-k}+ \vj^T_{m-l}\va
 \end{array}
 \right),
 \label{eq:U}
 \end{equation}
and is independent of the matrices $\vA$ and $\vB$.
The following corollary, that we refer to as \emph{Construction I}, is now immediate.
\smallskip

\begin{corollary}[{\em Construction I}]\label{cor:product}Let $\C$ and $\D$ be  linear $[n,k,d_\C]$ and $[m,l,d_\D]$ codes respectively
and $(\vI_k|\vA)$ and $(\vI_l|\vB)$ be their respective systematic encoders.
Pick $\vu=(\vzero_k,\va)\in (\FF_q^n/\C)_{\rm rep}$ and $\vv=(\vzero_l,\vb)\in (\FF_q^m/\D)_{\rm rep}$.
If in addition $\vj_n\in\C$ and $\vj_m\in\D$,
then  $(\C+\vu)\otimes (\D+\vv)$ defined by \eqref{eq:def1}
 is systematic of dimension $kl$ and is a coset of $\C\otimes \D$
with Property $(\C+\vu,\D+\vv)$.
\smallskip
\end{corollary} 

Binary linear codes that contain the all-one vector are called {\em self-complementary codes}.
Well-known examples of linear self-complementary codes 
include the primitive narrow-sense Bose-Chaudhuri-Hocquenghem codes,
the extended Golay code and 
the Reed-Muller codes \cite{MacWilliamsSloane:1977}.
Examples of $q$-ary linear codes that contain the all-one vector include
the Reed-Solomon codes, generalized Reed-Muller codes,
\cite{MacWilliamsSloane:1977}, and
difference matrix codes \cite{Bassalygo:2006a}.

\section{Variants of Construction I}
\label{sec:affine-example}

In this section, we adopt Construction I to 
certain nonlinear component codes $\C'$, $\D'$
that are variants of cosets of linear codes.
Several well-known families of nonlinear codes, such as Nordstrom-Robinson,
Delsarte-Goethals, Kerdock and Preparata, can be obtained as unions of cosets of linear codes
(see \cite[Ch. 15]{MacWilliamsSloane:1977}).
In general, it is difficult to achieve a
matrix code with Property $(\C',\D')$
of size  $q^{\log|\C'|\log |\D'|}$. Instead, we show that it is possible to achieve
a size of $q^{\kappa\log |\C'|\log |\D'|}$
for some positive constant $\kappa<1$.

A straightforward generalization of Construction I to 
union of cosets of linear codes can be achieved as follows.
Let $\C_1$ and $\D_1$ be  linear
$[n,k_1,d_{\C_1}]$ and $[m,l_1,d_{\D_1}]$ such that
$\vj_n\in\C_1$ and $\vj_m\in \D_1$.
Let $\U\subseteq (\FF_q^n/\C_1)_{\rm rep}$ and $\V\subseteq (\FF_q^m/\D_1)_{\rm rep}$.
We consider the component codes $\C'$ and $\D'$, where
\begin{equation*}
\C' =\bigcup_{\vu\in\U} \C_1+\vu, \mbox{ and }
\D' =\bigcup_{\vv\in\V} \D_1+\vv.
\end{equation*}
Then the \mbyn{} defined by
 \begin{equation}\label{def:union}
\bigcup_{\vu\in\U}\bigcup_{\vv\in\V} (\C_1+\vu)\otimes (\D_1+\vv).
\end{equation}
has Property $(\C',\D')$.
However, observe that  the code has size $$|\U||\V|q^{k_1l_1} = 
q^{k_1l_1+\log|\U|+\log|\V|},$$
while the sizes of $\C'$ and $\D'$ 
are $|\U|q^{k_1}=q^{k_1+\log|\U|}$ and $|\V|q^{l_1}=q^{l_1+\log|\V|}$ respectively.
Thus the size of the code obtained from \eqref{def:union} is
less than $q^{\log|\C'|\log|\D'|}=q^{(k_1+\log|\U|)(l_1+\log|\V|)}$.

\subsection{Product Construction of Expurgated Codes}

We improve the size given by \eqref{def:union} when 
the union of cosets of product codes has a certain structure.
Specifically, we consider the instance where the cosets form an expurgated code.
We describe this formally below.

In addition to the codes $\C_1, \D_1$,
 assume that $\C_2$ and $\D_2$ are  linear
 $[n,k_2,d_{\C_2}]$ and $[m,l_2,d_{\D_2}]$ codes such that
$\C_1\subset \C_2$ and $\D_1\subset \D_2$. 
We consider nonlinear component codes that are obtained from expurgated codes $\C_2\setminus \C_1$ and $\D_2 \setminus \D_1$. 
Our objective is therefore to construct an \mbyn{}
such that Property $(\C_2\setminus \C_1,\D_2\setminus \D_1)$ holds.

 Clearly, $\C_2\setminus \C_1$ and $\D_2\setminus \D_1$ are 
 union of cosets of $\C_1$ and $\D_1$
 with $\U=(\C_2/\C_1)_{\rm rep}\setminus\{ \vzero_n\}$ and
 $\V=(\D_2/\D_1)_{\rm rep}\setminus\{ \vzero_m\}$ respectively.
 Then the construction described in \eqref{def:union} gives a
 code with
size $(q^{k_2-k_1}-1)(q^{l_2-l_1}-1)q^{k_1l_1}\approx
q^{k_2-k_1+l_2-l_1+k_1l_1}$. The distance of the product code is determined
by the distance of the codes $\C_2$ and $\D_2$.

On the other hand, we improve this size via the following.
\medskip

\noindent{\bf Construction IA}. 
Consider two intermediary codes 
$\C_3$ and $\D_3$ of dimensions $k_2-1$ and $l_2-1$ respectively 
such that  $\C_1\subseteq \C_3 \subset \C_2$ and $\D_1\subseteq \D_3 \subset  \D_2$.
Pick any $\vu\in (\C_2\setminus \C_3)$ and $\vv\in (\D_2\setminus \D_3)$ and observe that
\begin{equation*}
\C_3+\vu \subset \C_2\setminus \C_1 \mbox{ and } \D_3+\vv \subset \D_2\setminus \D_1.
\end{equation*}

Applying Construction I to the cosets $\C_3+\vu$ and $\D_3+\vv$ yields 
a matrix code $(\C_3+\vu)\otimes(\D_3+\vv)$ with Property $(\C_3+\vu,\D_3+\vv)$,
and hence the Property $(\C_2\setminus \C_1,  \D_2\setminus \D_1)$.
Furthermore, the size of this code is $q^{(k_2-1)(l_2-1)}$ and 
is significantly larger than the straightforward construction from \eqref{def:union}.

\subsection{Binary Matrix Codes with Restricted Column and Row Weights}
\label{sec:plc}

 In this section, we apply Construction IA to obtain matrix codes with
 the additional property of bounded 
 row and column weights.
The motivation for studying such matrix codes arises from the application to
coded modulation for power line communication (PLC) channel.
Consider a codeword
$\vN \in \FF_2^{m\times n}$ of a matrix code. Each row of the
matrix corresponds to transmission over a particular frequency slot, while
each column of the matrix corresponds to a discrete time instance.
Transmision occurs at the frequency and time slots corresponding to a one in
the matrix. 

The different types of noises are as follows.
Assuming a hard-decision threshold detector, 
the received signal (which may contain errors caused by noise)
is demodulated to an output $\tN\in\FF_2^{m\times n}$.  
The burst and random errors that arise from the different types of
noises in the PLC channel (see \cite[pp. 222--223]{Ardakanietal:2010}) 
have the
following effects on the detector output. 
We denote the $(i,j)$-th entry of a matrix $\vN$
by $N_{i,j}.$
\begin{enumerate}
\item A narrowband noise introduces a tone at all time instances of the transmitted signals.
If $e\in [m] $ and $e$ {\em narrowband noise errors} occur, then there
is a set $\Gamma\in\binom{[m]}{e}$ of $e$ rows, such that $\widetilde N_{i,j}=1$ for $i\in \Gamma$, $j\in [n]$.

\item Impulse noise results in the entire set of tones being received at a certain time instance.
If $e\in [n]$ and $e$ {\em impulse noise errors} occur, 
then there is a set $\Pi\in\binom{[n]}{e}$
of $e$ columns such that $\widetilde N_{i,j}=1$ for $i\in [m]$, $j\in \Pi$.
\item A channel fade event erases a particular
    tone. If $e\in[m]$, and $e$ {\em fades} occur then there is a set
    $\Gamma \in \binom {[m]}e$ of $e$ rows such that $\widetilde N_{i,j} = 0$ for all
    $j\in[n]$.
\item Background noise flips the value of the bit at a particular tone
    and time instance. If $e$ {\em background noise} occurs then there exists
    a set $\Omega \in \binom{[n]\times[m]}e$ such that $\widetilde N_{i,j}
    = N_{i,j}+1,$ for all $(i,j)\in\Omega.$
\end{enumerate}
We refer to \cite{Ardakanietal:2010} for an
expanded description of the types of noise that are present in the power
line channel.

If any row of the codeword matrix $\vN$ is an
all-one vector then this row is not distinguishable from an all-one row
introduced by the presence of narrowband noise. Similarly, an all-one column is not
distinguishable from impulse noise.
Additionally, the use of multi-tone frequency shift keying is adopted with
the understanding that the energy is concentrated on a fraction of
the available frequencies (see \cite{Cheeetal:2013c}). 
Thus, it is desired that every row and
every column of the matrix contain at least a single one, but it should not
be an all-one vector.
This requires the use of codes whose codewords are
matrices with restricted and
bounded column and row weights.

In particular, for the application to powerline communications we construct
codes which are able to correct narrowband and impulse noise,
random noise, and also simultaneously satisfy all of the following criteria
(also see \cite{Cheeetal:2013c}):
\begin{enumerate}
    \item[(A1)] have positive rate,
    \item[(A2)] have positive relative distance,
    \item[(A3)] have efficient decoding algorithms, and
    \item[(A4)] have no restriction that the length of the code is at most the
        size of the alphabet.
\end{enumerate}
\medskip
 In the following text, we use the following lemma that was crucial in proving the so-called low symbol weight property (see
 \cite[Proposition 1]{Versfeldetal:2010}) for $q$-ary affine codes.
\medskip

 \begin{lemma}\label{lem:dnd}
 Let $\C$ be binary linear $[n,k,d]$  code such that
 $\inprod{\vj_n}\subset \C$.
 Then the codewords in $\C \setminus \inprod{\vj_n}$ have Hamming weight
 bounded between $d$ and $n-d$.
 \end{lemma}
\smallskip

 \begin{proof}
    If the Hamming weight of any vector $\vv$ is greater than $n-d$, then
    $\vv + \vj_n$ has Hamming weight less than $d$. This is
    a contradiction.
 \end{proof}
\medskip
 
First, we illustrate via an example that the code obtained by
straightforward expurgation does not satisfy the systematic property.
\medskip

\begin{example}
    \label{exa:90}
 Let $\C=\D$ be the binary linear $[4,3,2]$ code consisting of all even weight codewords.
 Observe that $\C\setminus \inprod{\vj_4}$ consists of six codewords of weight two
and we are interested in constructing a $(4\times 4)$-matrix code whose matrices have  row weight two and column weight two.

A naive approach is to look at the $(3\times 3)$ information matrix and 
require all columns and rows to not belong to $\{\vzero_3,\vj_3\}$.
This approach fails as illustrated by the example codeword,
\begin{equation*}
\left(
      \begin{array}{ccc|c}
      1 & 0 & 0 & 1\\
      0 & 1 & 0 & 1\\
      0 & 0 & 1 & 1\\ \hline
      1 & 1 & 1 & 1
      \end{array}
      \right),
\end{equation*}
which contains  an all-one row even though each of the component codewords in
the first three rows and columns have weight exactly two.

On the other hand, consider the binary linear $[4,2,2]$ code
$\C_3=\{\vzero_4,\vj_4, (1,0,1,0),$ $ (0,1,0,1)\}$ and let $\vu=(0,0,1,1)$.
Then $(\C_3+\vu)\otimes (\C_3+\vu)$ yields a
$(4\times 4)$-matrix code whose matrices have row weight two and column weight two.
Furthermore, it is systematic of dimension four.

On the other hand, it can be obtained via
 computer search that there are exactly $90$ matrices in $\C\otimes \C$  
that have constant row weight two and constant column weight two.
An exhaustive computer search shows that there do not exist
five coordinates where a subset of these $90$ matrices is systematic.
\end{example}
\medskip

We proceed with the construction of matrix codes with restricted row and
column weights.
Let $\C$, $\D$ be binary linear 
$[n,k,d_{\C}]$, and $[m,l,d_{\D}]$ codes respectively.
Suppose 
$\inprod{\vj_n} \subset \C$, and 
$\inprod{\vj_m} \subset \D$.
Direct application of Construction IA yields
a systematic binary \mbyn{} of dimension $(k-1)(l-1)$
whose matrices have 
\begin{enumerate}[(i)]
\item row weight bounded between $d_{\C}$ and $n-d_{\C}$, 
\item column weight bounded between $d_{\D}$ and $m-d_{\D}$.
\end{enumerate}

In Example \ref{exa:90} we showed that this construction gives more
desirable results and why naive methods of constructions do not work.
Because of the narrowband and impulse noise present in the power line
channel, we want codes with restricted column and row weights. The
following proposition gives the condition under which the noises can be
corrected.
\medskip

\begin{proposition}\label{prop:nbnoise}
Let $\C$, $\D$ be binary linear 
$[n,k,d_{\C}]$, and $[m,l,d_{\D}]$ codes respectively.
Suppose 
$\inprod{\vj_n} \subset \C$, and 
$\inprod{\vj_m} \subset \D$.
Then $(\C\setminus \inprod{\vj_n})\otimes (\D\setminus \inprod{\vj_m})$
obtained using Construction IA yields
a systematic binary \mbyn{} of dimension $(k-1)(l-1)$
whose matrices have 
 \begin{enumerate}[(i)]
 \item row weight bounded between $d_{\C}$ and $n-d_{\C}$, 
  \item column weight bounded between $d_{\D}$ and $m-d_{\D}$.
\end{enumerate}

\noindent Furthermore,  $(\C\setminus \inprod{\vj_n})\otimes (\D\setminus
\inprod{\vj_m})$
 is a subcode of $\C\otimes \D$,
and hence, is able to correct $\enarrow$ narrowband errors and $\eimpulse$ impulse noise errors, 
provided
\begin{equation*}
\eimpulse < d_{\C}, \mbox{ and }
\enarrow < d_{\D}.
\end{equation*}
 \end{proposition}
\smallskip

\begin{proof}
    Consider a code $\C' \subset \C$ of
    dimension $k-1$, and $\D'\subset \D$ of dimension $l-1$. Using
    Construction IA, we consider the cosets $\C'+\vu$, and $\D'+\vv$
    where $\vu \in \C\setminus\C'$ and $\vv \in \D\setminus\D'.$ By Lemma
    \ref{lem:dnd}, the weight of every vector in $\C'+\vu$ is bounded
    between $d_\C$ and $n - d_{\C}$. Similarly, condition (ii) holds.

    To correct $\enarrow$ narrowband noise errors and $\eimpulse$ impulse
    noise errors, we use Algorithm \ref{nberror} which first sets all
    narrowband noise and impulse noise errors to erasures,
    subsequently subtracts the
    coset leader, and then decodes the row and the column codes.
    In the absence of random errors,
    if the condition $\eimpulse < d_{\C}$ is satisfied, then the row code
    $\C'$ can correct all the corresponding erasures. Similarly, the column
    code $\D'$ can correct  all the erasures in each column if
    $\enarrow < d_{\D}$.
\end{proof}
\medskip
\begin{algorithm}[!hbt]
\SetAlgoLined
\KwIn{detector output $\tN\in \FF_2^{m\times n}$, coset leader $\vU$} 
\KwOut{$\vN'\in \C'\circ\D'$} 
\BlankLine
\tcc{Consider the narrowband noise as erasures}
\For{$i\in [m]$}{
    \If{$\tN_{i,j}=1$ for all $j\in[n]$}
    {$\tN_{i,j} \gets \varepsilon$ for all $j\in[n]$}
}
\BlankLine
\tcc{Consider the impulse noise as erasures}
\For{$j\in[n]$}{
    \If{$\tN_{i,j}\in\{1,\varepsilon\}$ for all $i\in[m]$}{
        $\tN_{i,j}\gets\varepsilon$ for all $i\in[m]$
    }
}
\BlankLine
\tcc{Subtract the coset leader from the nonerased coordinates}
\For{$i\in [m], j\in[n]$}{
    \If{$\tN_{i,j}\ne\varepsilon$}{
        $\tN_{i,j}\gets\tN_{i,j}-\vU_{i,j}$
    }
}
Decode $\tN$ to $\vN'$ using a product code decoder\\
\Return{$\vN'$}
\caption{Decoder for Product of Affine Codes}
\label{nberror}
\end{algorithm}
\medskip

\subsubsection{Optimality of the product construction}
The affine codes obtained using the product construction Construction IA,
are likely not optimal for large dimensions of the matrix. Obtaining
optimal codes which satisfy all the criteria (A1)--(A4) stated earlier in
this subsection is still an open problem. Below, we show some examples of
codes for which the construction is close to optimal.
\medskip

\begin{example}
    \label{exa:reedmuller}
    Consider the first order Reed-Muller code with parameters $\C = [2^r, r+1,
    2^{r-1}]$. The affine code $(\C \setminus \inprod{\vj_n})\otimes
    (\C\setminus \inprod{\vj_n})$ obtained by Construction IA is a $(2^r
    \times 2^r)$-matrix code of dimension $r^2$ and Hamming distance
    $2^{2r-2}.$ This code can correct $\eimpulse < 2^{r-1}$ impulse noise
    errors and $\enarrow < 2^{r-1}$  narrowband noise errors.
\end{example}
\medskip

Before providing the next example, we recall a ``Gabidulin Construction''
from the thesis of the second author \cite[Section 5.4]{kiah:2013} for square matrix codes.
\medskip

\begin{proposition}[Kiah \cite{kiah:2013}]
    \label{prop:gabidulin}
    Let $d < n$ and let $k = n-d+1.$ Then there exists a binary $(2n \times
    2n)$ matrix code of size $2^{nk}$ with constant column weight $n$ that
    corrects $\eimpulse$ impulse noise errors and $\enarrow$ narrowband
    noise errors provided that $\eimpulse < n$, $\enarrow < n$, and
    $$
    \floor{\frac{\eimpulse}2} + \floor{\frac{\enarrow}2} < d.
    $$
\end{proposition}
The family of codes in the above proposition is obtained from Gabidulin
codes \cite{Gabidulin:2008}, which are optimal rank metric codes,
and can be explicitly written as follows. Let $\C$ denote a Gabidulin code,
and let $\C^*$ denote the matrix code obtained using Proposition
\ref{prop:gabidulin}.
Then we get,
$$
\C^* = \left\{ \begin{pmatrix}\vM & \vM+\vJ\\ \vM+\vJ & \vM\end{pmatrix}: \vM \in
\C\right\},
$$
where $J$ is the all-one matrix. We now proceed to provide an example
similar to Example \ref{exa:reedmuller}.
\medskip

\begin{example}
    \label{exa:gabidulin}
    Consider the Gabidulin code with parameters $[n = 2^{r-1}, 1,
    d = 2^{r-1}]$. Such a code has dimension $nk = 2^{r-1}$, and can
    correct the same number of narrowband and impulse noise errors that the
    code in Example \ref{exa:reedmuller} can correct. 
    We get the following table comparing the dimensions of the two codes
    obtained from Example \ref{exa:reedmuller} and Proposition
    \ref{prop:gabidulin} respectively.
    \smallskip
    
\begin{center}
    \begin{tabular}{c|c|c}
    		& Dimension of codes from  & Dimension of codes from \\
        $r$ & Example \ref{exa:reedmuller}, $r^2$ &  Proposition \ref{prop:gabidulin}, $2^{r-1}$ \\
        \hline
        $3$   & $9 $    &   $4 $      \\
        $4$   & $16$    &   $8 $      \\
        $5$   & $25$    &   $16$      \\
        $6$   & $36$    &   $32$      \\
        $7$   & $49$    &   $64$ 
    \end{tabular}
        \end{center}
        \smallskip

    Thus, beyond $r=6$, the construction from the Gabidulin codes has
    better parameters than the product construction. However, it must be
    noted that in the case of the Gabidulin construction, we have
    restricted the matrices in our matrix code to always be a square
    matrix.
\end{example}

 For fair comparison, we have taken the affine codes to also be formed of
square matrices with the same number of rows and columns as in the
Gabidulin construction.  The product code
in general does not have this restriction.  This satisfies point (A4) in
the criteria that we want to satisfy. Therefore, it is of interest (and an
open problem) to obtain the true upper bound for codes which are optimal
and which satisfy all the four criteria (A1)--(A4).

\section{Irregular Product of Affine Codes}
\label{sec:irregular}
The power line channel is known to be frequency selective (see
\cite{Ardakanietal:2010}), i.e., the background noise in different
frequency slots are of different intensities. Thus, it is of interest to
provide constructions of codes that can provide different levels of error
correction over different frequencies. Such codes can be constructed as
product codes where the rows of the matrix correspond to different row
codes. Such codes have been studied earlier as ``generalized concatenated
codes'' or ``multilevel concatenated codes'' (see Blokh and Zyablov
\cite{Zyablov:1974}, Zinoviev \cite{Zinoviev:1976}, and Dumer \cite{Dumer:1998}).
The row codes,
which correspond to the row encoding, in these constructions are defined
over an extension field of the field of the column code. As a result,
although the resulting matrix code is linear over the smaller field, the
rows do not in general belong to the row code. This makes it difficult to
extend the construction to product of affine codes.

Instead, we consider the case where the component codes for each row and
column are different. 
Although the application is only for binary component codes, we provide the
general theory for $q$-ary component codes.
Such product codes were termed {\em irregular
product codes} and were studied by Alipour \etal \cite{Alipour:2012}.
Specifically, they demonstrated the following proposition.
%
\medskip

\begin{proposition}[Alipour \etal \cite{Alipour:2012}]\label{prop:irregular}
Let  $\C_i$ be a linear code of length $n$ and dimension $k_i$ for  $i\in[m]$ and  
$\D_j$ be a linear code of length $n$ and dimension $l_j$ for  $j\in[m]$.
Suppose that $k_1\le k_2\le \cdots \le k_m$ and $l_1\le l_2\le \cdots \le l_n$.
Then there exists a linear  \mbyn{} of dimension
$K$, where
\begin{equation}\label{irregular}
K\le \sum_{j=1}^n \sum_{i=l_{j-1}+1}^{l_j} \max\{k_i-j+1,0\}, \text{ where } l_0=0,
\end{equation}
and every codeword $\vN$ satisfies the properties that
\begin{enumerate}[(i)]
\item the $i$-th row of $\vN$ belongs to $\C_i$ for $i\in [m]$, and
\item the $j$-th column of $\vN$ belongs to $\D_j$ for $j\in [n]$.
\end{enumerate}
Furthermore, if 
$\C_1\subseteq\C_2\subseteq\cdots\subseteq \C_m$ and 
$\D_1\subseteq\D_2\subseteq\cdots\subseteq \D_n$,
we achieve equality in \eqref{irregular}.
\end{proposition}
\medskip

The encoding algorithm of irregular product codes is described in
\cite{Alipour:2012}. The encoding procedure encodes the rows first and then
encodes the columns. The encoding assumes that the first $k_i$ coordinates
of the $i$-th row can generate the remaining $n-k_i$ coordinates of that
row, and that the first $l_j$ coordinates of the $j$-th column can generate
the remaining $m-l_j$ coordinates of that column. Since the generating
coordinates of the code are present within the leading principal $l_n
\times k_m$ submatrix, we have the following Lemma.
\medskip

\begin{lemma}
    \label{lem:lnkm}
    The leading principal $l_n\times k_m$ submatrix generates all the
remaining coordinates of a codeword $\vN$ in the irregular product code.
\end{lemma}
\smallskip


We apply Construction I directly to Proposition \ref{prop:irregular} to
obtain an irregular product of affine codes.
\medskip

\begin{proposition}\label{prop:irregularaffine}
    In addition to the conditions of Prop. \ref{prop:irregular},
let $\vj_n \in C_i$ for $i\in [m]$ and
$\vj_m \in D_j$ for $j\in [n]$.
Let $\vu=(\vzero_{k_m},\va)\in\FF_q^n\setminus \bigcup_{i=1}^m \C_i$ and
$\vv=(\vzero_{l_n},\vb)\in\FF_q^m\setminus \bigcup_{j=1}^n \D_j$.

Then there exists an affine  \mbyn{} of dimension $K$ 
bounded by \eqref{irregular}
and every codeword $\vN$ in the code satisfies the properties that
\begin{enumerate}[(i)]
\item the $i$-th row of $\vN$ belongs to $\C_i+\vu$ for $i\in [m]$, and 
\item the $j$-th column of $\vN$ belongs to $\D_j+\vv$ for $j\in [n]$.
\end{enumerate}
If $\C_1\subseteq\C_2\subseteq\cdots\subseteq \C_m$ and
$\D_1\subseteq\D_2\subseteq\cdots\subseteq \D_n$, we achieve equality in
\eqref{irregular}.
\smallskip

For $q=2$, suppose there exist linear codes $\C$ and $\D$ such that
$\bigcup_{i=1}^m \C_i \subset \C$ and 
$\bigcup_{i=1}^n \D_i \subset \D$, respectively.
For $\vu\in\C\setminus \bigcup_{i=1}^m \C_i$ and
$\vv\in\D\setminus \bigcup_{j=1}^n \D_j$,
the weight of every
row of any codeword is bounded between $d_\C$ and $n - d_\C$, and of every
column between $d_\D$ and $m-d_\D$, where $d_\C$ and $d_\D$ are the minimum
distances of $\C$ and $\D$ respectively.
\end{proposition}
\medskip

\begin{proof}
    Let $\vN$ be a codeword obtained by using the encoding described in
    \cite{Alipour:2012}.
We translate this codeword using the matrix
 \begin{equation*}
 \vU=\left(
 \renewcommand\arraystretch{2} 
 \begin{array}{c|c}
     {\vzero_{k_m\times l_n}} & \vj^T_{l_n}\va\\ \hline
     \vb^T\vj_{k_m} & \vb^T\vj_{n-k_m}+ \vj^T_{m-l_n}\va
 \end{array}
 \right),
 \end{equation*}
where the vector $\va$ has length $n - k_m$ and $\vb$ has length $m-l_n$.
We denote the codeword $\vN$ by four submatrices, as
$$
\vN = \begin{pmatrix}\vN_1 & \vN_2\\ \vN_3 & \vN_4\end{pmatrix},
$$
where $\vN_1$ is the $l_n \times k_m$ leading principal submatrix that
generates $\vN_2, \vN_3, \vN_4$, by Lemma \ref{lem:lnkm}. The submatrix
$\vN_2$  is of dimension $l_n \times (n-k_m)$,
$\vN_3$ is of dimension $(m-l_n)\times k_m$, and $\vN_4$ is of dimension
$(m-l_n)\times(n-k_m).$ Denote the corresponding matrix from the coset code
as
$$
\vN' = \begin{pmatrix}\vN_1 & \vN_2'\\ \vN_3' & \vN_4'\end{pmatrix},
$$
where $\vN_2' = \vN_2 + \vj_{l_n}^T \va$, and $\vN_3' = \vN_3 + \vb^T
\vj_{k_m}$.

We need to ensure that the matrix $\vN_4'$ obtained by encoding the rows
of $\vN_3'$ by the row codes satisfies the condition that they belong to
the row code, and also satisfies that they belong to the
column codes in the submatrix $\begin{pmatrix}\vN_2'\\
\vN_4'\end{pmatrix}.$ This can be proved as follows.
Let the generator matrices
of $\C_i,\ i=l_n+1,\dots,m$ be given by the matrices $G_i = [I_{k_i}| A_i'
| A_i]$, where $A_i'$ has dimension $k_i \times (k_m - k_i)$ and $A_i$ has
dimension $k_i \times (n - k_m)$.
Let $\vb = (b_{l_n+1},\dots,b_m)$.
The $i$-th row of $\vN_3'$ can be split into two parts, corresponding to
the first two blocks of the generator matrix $G_i$ as follows. For
$i=l_n+1,\dots,m$ we first
write
the $i$-th row of $\vN_3$ as $\vN_{3,i} = (\vn_i,\vn_i')$ where the first block
has length $k_i$ and the second block has length $k_m - k_i.$ We obtain,
$$
\vN_{3,i}' = (\vn_i + b_i \vj_{k_i}, \vn_i' + b_i \vj_{k_m - k_i})
= \vN_{3,i} + b_i \vj_{k_m}.
$$
Encoding the first block of this row $\vN_{3,i}'$ with the generator matrix
$G_i$ gives the vector
\begin{align*}
    (\vn_i + b_i \vj_{k_i}) [I_{k_i}|A_i'|A_i] &=
(\vn_i + b_i \vj_{k_i}, \vn_iA_i' + b_i \vj_{k_m - k_i},
\vn_i A_i + b_i \vj_{n - k_m})\\
 & = (\vn_i, \vn_i', \vN_{4,i}) + b_i \vj_n
\end{align*}
where $\vn_i A_i' = \vn_i'$ is the second block of $\vN_{3,i}$,
and $\vn_iA_i = \vN_{4,i}$ is the $i$-th row of $\vN_4$. The shift by the
coset leader $(\vzero_{k_m}, \va)$ results in the word
$$((\vn_i, \vn_i') + b_i \vj_{k_m},\, \vN_{4,i} + b_i \vj_{n-k_m} + \va)
= (\vN_{3,i}', \vN_{4,i} + b_i \vj_{n-k_m} + \va).$$
Thus, the matrix $\vN_4'$ is given by the expression
$$
\vN_4' = \vN_4 + \vb^T \vj_{n - k_m} + \vj_{m - l_n}^T \va.
$$
A similar argument shows that the submatrix $\begin{pmatrix}\vN_2'\\ \vN_4'
\end{pmatrix}$ satisfies the corresponding column codes.

The same argument
as in the proof of Proposition \ref{prop:nbnoise} shows that the row and
column weights are bounded when the conditions $\C_1\subseteq \C_2 \subseteq
\cdots \subseteq \D_m \subset \C$, and $\D_1\subseteq \D_2 \subseteq
\cdots \subseteq \D_m \subset \D$ hold.
\end{proof}

\section{Conclusion}
We provide new constructions of systematic nonlinear product codes that are
obtained by taking product of cosets of linear codes. The constructions
have the property that every row and every column belongs to the row code
and column code, respectively. Subsequently, we show that it is possible to
construct matrix codes with restricted column and row weights. Although the
primary motivation for studying such matrix codes is for coded modulation
over power line channel, the constructions can potentially be adapted to
address other problems where such codes are desired such as codes for
memristor arrays and two-dimensional weight-constrained codes
\cite{Ordentlich:2011,Ordentlich:2000}.

\section*{Acknowledgement}
The authors thank the referee for the useful comments, which helped clarify
the presentation.

This work was done while P.~Purkayastha was
a Research Fellow at Nanyang Technological University, Singapore.

\bibliographystyle{siam}

\begin{thebibliography}{10}

\bibitem{Alipour:2012}
{\sc M.~Alipour, O.~Etesami, G.~Maatouk, and A.~Shokrollahi}, {\em Irregular
  product codes}, in IEEE Information Theory Workshop (ITW) 2012, 2012,
  pp.~197--201.

\bibitem{Amrani:2007}
{\sc O.~Amrani}, {\em Nonlinear codes: The product construction}, IEEE Trans.
  Commun., 55 (2007), pp.~1845--1851.

\bibitem{Ardakanietal:2010}
{\sc M.~Ardakani, G.~Colavolpe, K.~Dostert, H.~C. Ferreira, D.~Fertonani, T.~G.
  Swart, A.~M. Tonello, D.~Umehara, and A.~J.~H. Vinck}, {\em Digital
  transmission techniques}, in Power Line Commun. - Theory and Applicat. for
  Narrowband and Broadband Commun. over Power Lines, H.C. Ferreira, L.~Lampe,
  J.~Newbury, and T.G. Swart, eds., John Wiley \& Sons, 2010, ch.~5,
  pp.~195--310.

\bibitem{Barg:2011}
    {\sc A.~Barg and G.~Z{\'e}mor}, {\em List decoding of product codes by the minsum
  algorithm}, in IEEE International Symposium on Information Theory Proceedings
  (ISIT), 2011, pp.~1273--1277.

\bibitem{Bassalygo:2006a}
{\sc L.~A. Bassalygo, S.~M. Dodunekov, V.~A. Zinoviev, and T.~Helleseth}, {\em
  The Grey-Rankin bound for non-binary codes (russian)}, Problemy Pereda\v ci
  Informacii, 42 (2006), pp.~37--44; translation in Probl. Inform. Trans., 42
  (2006), pp. 197--203.

\bibitem{Zyablov:1974}
{\sc \`E.~L. Blokh and V.~V. Zyablov}, {\em Coding of generalized concatenated
  codes}, Problemy Pereda\v ci Informacii, 10 (1974), pp.~45--50.

\bibitem{Cheeetal:2013c}
{\sc Y.~M. Chee, H.M. Kiah, and P.~Purkayastha}, {\em Matrix codes and
  multitone frequency shift keying for power line communications}, in Proc.
  IEEE Intl. Symp. Inform. Theory, 2013, pp.~2870--2874.

\bibitem{Cheeetal:2014}
{\sc Y.~M. Chee, H.M. Kiah, P.~Purkayastha, and P.~Sol\'e}, {\em Product
  construction of affine codes}, in Proc. IEEE Intl. Symp. Inform. Theory,
  Honolulu, USA, 2014, pp.~1441--1445.

\bibitem{Dumer:1998}
{\sc I.~Dumer}, {\em Concatenated codes and their multilevel generalizations},
  Handbook of coding theory, 2 (1998), pp.~1911--1988.

\bibitem{Elias:1954}
{\sc P.~Elias}, {\em Error-free coding}, Transactions of the IRE Professional
  Group on Information Theory, 4 (1954), pp.~29--37.

\bibitem{Forney:1966}
{\sc G.D. Forney~Jr}, {\em Concatenated codes. research monograph no. 37},
  1966.

\bibitem{Gabidulin:2008}
{\sc Ernst~M Gabidulin and Nina~I Pilipchuk}, {\em Error and erasure correcting
  algorithms for rank codes}, Designs, codes and Cryptography, 49 (2008),
  pp.~105--122.

\bibitem{kiah:2013}
{\sc H.~M. Kiah}, {\em Reliable Communications over Power Lines through Coded
  Modulation Schemes}, PhD thesis, Nanyang Technological University, Singapore,
  2013.

\bibitem{MacWilliamsSloane:1977}
{\sc F.~J. MacWilliams and N.~J.~A. Sloane}, {\em The Theory of
  Error-Correcting Codes}, North-Holland Publishing Co., Amsterdam, 1977.

\bibitem{Ordentlich:2000}
{\sc E.~Ordentlich and R.M. Roth}, {\em Two-dimensional weight-constrained
  codes through enumeration bounds}, IEEE Transactions on Information Theory,
  46 (2000), pp.~1292--1301.

\bibitem{Ordentlich:2011}
\leavevmode\vrule height 2pt depth -1.6pt width 23pt, {\em Low complexity
  two-dimensional weight-constrained codes}, in IEEE International Symposium on
  Information Theory Proceedings (ISIT) 2011, 2011, pp.~149--153.

\bibitem{Versfeldetal:2010}
{\sc D.~J.~J. Versfeld, A.~J.~H. Vinck, J.~N. Ridley, and H.~C. Ferreira}, {\em
  Constructing coset codes with optimal same-symbol weight for detecting
  narrowband interference in ${M}$-{FSK} systems}, IEEE Trans. Inform. Theory,
  56 (2010), pp.~6347--6353.

\bibitem{Vinck:2000}
{\sc A.~J.~H. Vinck}, {\em Coded modulation for power line communications},
  AE\"{U} - Int J. Electron. Commun., 54 (2000), pp.~45--49.

\bibitem{Zinoviev:1976}
{\sc V.~A. Zinoviev}, {\em Generalized cascade codes}, Problemy Pereda\v ci
  Informacii, 12 (1976), pp.~5--15.

\end{thebibliography}

\end{document}